\documentstyle[12pt]{article}

\catcode`@=11 \@addtoreset{equation}{section} \catcode`@=12

\begin{document}
\begin{titlepage}
\begin{flushright}\vbox{\begin{tabular}{c}
           TIFR/TH/97-08\\
           March, 1997\\
           hep-ph/9703370\\
\end{tabular}}\end{flushright}
\begin{center}
   {\large \bf
      Higher Orders in the Colour-Octet Model\\
      of $J/\psi$ Production}
\end{center}
\bigskip
\begin{center}
   {Sourendu Gupta\footnote{E-mail: sgupta@theory.tifr.res.in}
    and Prakash Mathews\footnote{E-mail: prakash@theory.tifr.res.in}.\\
    Theory Group, Tata Institute of Fundamental Research,\\
    Homi Bhabha Road, Bombay 400005, India.}
\end{center}
\bigskip
\begin{abstract}
We study the hadro- and photo-production of $\bar cc$ and $\bar bb$ mesons
at low transverse momentum to high orders in the relative velocity of the
pair, $v$, in non-relativistic QCD. We evaluate cross sections to order
$v^7$ for $\eta_c$, sufficient for studies of photo-production in the
almost-elastic region. For all other charmonium states we find the cross
section to order $v^9$, sufficient for studies of the ratio of $\chi_{c1}$
and $\chi_{c2}$ production rates. We find recurrence formul\ae{} for
generating terms at even higher orders, should they be needed.
\end{abstract}
\end{titlepage}

\section{\label{intro}Introduction}

Recent progress in the understanding of cross sections for production of
heavy quarkonium resonances has come through the non-relativistic QCD (NRQCD)
reformulation of this problem \cite{caswell}. Although the production of the
pair is dominated by short distance scales of order $1/m$ (where $m$ is the
heavy quark mass), longer non-perturbative scales play an important role.
The projection on to a specific quarkonium state involves length scales
such as $1/mv$, $1/mv^2$, {\sl etc.\/}, where $v$ is the (dimensionless)
velocity of either of the heavy fermions in the rest frame of the pair.

Factorisation of the physics at the long and short scales has been proven
in the NRQCD formalism for processes dominated by a large transverse momentum
\cite{bbl}. The resulting cross sections are a double power series in the
QCD coupling $\alpha_{\scriptscriptstyle S}$ evaluated at the NRQCD
factorisation scale $\mu_0$, and the velocity $v$. Often, higher orders in
$v$ involve the previously neglected colour-octet states of the heavy quark
pairs. For bottomonium states, $v^2\ll\alpha_{\scriptscriptstyle S}(m^2)$
and hence colour octet contributions are often not very significant and
the expansion is close to the normal perturbative expansion. For charmonium
states, a numerical coincidence, $v^2\sim\alpha_{\scriptscriptstyle S}(m^2)$,
makes the double expansion more complicated.

The formalism has been successfully applied to large transverse momentum
processes \cite{jpsi}. Interestingly, inclusive production cross sections
for charmonium at low energies, dominated by low transverse momenta, also
seem to have a good phenomenological description in terms of this approach
\cite{ours,br,our2}. It was argued \cite{our2} that a better understanding
of such cross sections can be obtained if the higher order terms in $v$ and
$\alpha_{\scriptscriptstyle S}$ are used. The argument is simple. Total
inclusive $J/\psi$ cross sections arise either from direct $J/\psi$
production (which starts at order $\alpha_{\scriptscriptstyle S}v^7$) or
through decays of $\chi$ states. Now $\chi_0$ and $\chi_2$ are first produced
at order $\alpha_{\scriptscriptstyle S}v^5$, whereas $\chi_1$, which has the
largest branching fraction into $J/\psi$, is produced only at order
$\alpha_{\scriptscriptstyle S}v^9$. Hence, a full understanding of these
cross sections requires the NRQCD expansion upto order
$\alpha_{\scriptscriptstyle S}v^9$. We calculate these terms here. Of course,
higher orders in $\alpha_{\scriptscriptstyle S}$ at lower orders in $v$ may
be equally important. First attempts at computing these have been made
\cite{pert}.

The plan of this paper is the following. In section 2 we set out the notation
with a brief review of the threshold expansion technique \cite{bchen}, and
some extensions, which we shall use for our computation. In the
next section we perform a Taylor expansion of the transition matrix
elements in the relative momentum of the heavy quark pair and give
recurrence relations for the Taylor coefficients to all orders. The cross
sections are listed in section 4. We conclude in section 5 with a discussion
of the phenomenological applications of this computation. The appendix
contains a discussion of the complete specification of states and operators
required for computations beyond the leading orders of NRQCD.

\section{The Threshold Expansion}

The NRQCD factorisation formula for inclusive production of heavy quarkonium
resonances $H$ with 4-momentum $P$ is
\begin{equation}
   d\sigma\;=\;{1\over\Phi}{d^3P\over(2\pi)^3 2E_{\scriptscriptstyle P}}
     \sum_{ij} C_{ij}\left\langle{\cal K}_i\Pi(H){\cal K}^\dagger_j\right\rangle,
\label{te.nrqcd}\end{equation}
where $\Phi$ is a flux factor. The coefficient function $C_{ij}$ is
computable in perturbative QCD and hence has an expansion in the strong
coupling $\alpha_{\scriptscriptstyle S}$ (evaluated at the NRQCD cutoff),
whereas the matrix element is non-perturbative. However, in NRQCD, it has
a fixed scaling dimension in the quark velocity $v$. Consequently, the
cross section is a double power series in $\alpha_{\scriptscriptstyle S}$
and $v$.

The fermion bilinear operators ${\cal K}_i$ are built out of heavy quark
fields sandwiching colour and spin matrices and the covariant derivative
${\bf D}$. The specification of the composite labels $i$ and $j$ is given
in appendix \ref{coupl}. They include the colour index $\alpha$, the spin
quantum number $S$, the orbital angular momentum $L$ (found by coupling
the derivatives), the total angular momentum $J$ and the helicity $J_z$.
At low orders in $v$ this set is sufficient to fix the operators. Since
the hadronic projection operator 
\begin{equation}
   \Pi(H)\;=\;{\sum_s} \left|H,s\rangle\langle H,s\right|,
\label{te.hproj}\end{equation}
(where $s$ denotes hadronic states with energy less than the NRQCD cutoff),
is diagonal in these quantum numbers, it is clear that the operators ${\cal
K}_i$ and ${\cal K}_j$ in eq.\ (\ref{te.nrqcd}) are restricted to have equal
$L$, $S$, $J$ and $J_z$. For a more detailed discussion see the appendix.

The $J_z$-dependence of these matrix elements can be factored out using the
Wigner-Eckart theorem---
\begin{equation}
   \langle{\cal K}_i\Pi(H){\cal K}^\dagger_i\rangle\;=\;
		   {1\over2J+1} {\cal O}^H_\alpha({}^{2S+1}L_J^N),
\label{we.ored}\end{equation}
where the first factor on the right comes from a Clebsch-Gordan coefficient.
This factor is conventionally included in the coefficient function.
In the reduced matrix element ${\cal O}^H$, we have introduced a new label
$N$ which is the number of derivatives in each fermion bilinear (see the
appendix). In agreement with the notation of \cite{bbl} we write for the
off-diagonal operators
\begin{equation}
   \langle{\cal K}_i\Pi(H){\cal K}^\dagger_j\rangle\;=\; {1\over2J+1}
            {\cal P}^H_\alpha({}^{2S+1}L_J^N,{}^{2S+1}L_J^{N'}).
\label{we.pred}\end{equation}
The power counting rule for the matrix
elements in eq.\ (\ref{te.nrqcd}) is---
\begin{equation}
   d\;=\;3+N+N'+2(E_d+2M_d),
\label{te.rule}\end{equation}
where $E_d$ and $M_d$ are the number of colour electric and magnetic
transitions required to connect the hadronic state to the state
${\cal K}_i|0\rangle$. Note that at low orders in ${\bf D}$, $N=L$,
and the more familiar rules are obtained. An example is provided by
the off-diagonal matrix element
\begin{equation}
  {\cal P}^{\eta_c}_1\left({}^1S_0^0,{}^1S_0^2\right) \;\equiv\;
    -{1\over\sqrt3}
    \left\langle\psi^\dagger\chi\Pi(\eta_c)\chi^\dagger
     (-{i\over2}{\bf D})\cdot(-{i\over2}{\bf D})\chi\right\rangle
    +{\rm h.c.}
\label{pc.exam}\end{equation}
which scales as $v^5$. The $-1/\sqrt3$ factor on the right is a trivial
Clebsch-Gordan factor, explained later.

We choose to construct the coefficient functions using the ``threshold
expansion'' technique of \cite{bchen}. This consists of calculating, in
perturbative QCD, the matrix element ${\cal M}$ connecting the initial
states to final states with a heavy quark-antiquark pair ($\bar QQ$),
and Taylor expanding the result in the relative momentum of the pair,
$q$, after performing a non-relativistic reduction of the Dirac spinors.
The resulting expression is squared and matched to the NRQCD formula of
eq.\ (\ref{te.nrqcd}) by inserting a perturbative projector onto a
non-relativistic $\bar QQ$ state between the two spinor bilinears. The
coefficient of this matrix element is the required coefficient function.

Symbolically---
\begin{equation}
   \overline{\sum_{pol}} |{\cal M}|^2 \;=\;
     \sum_{ij} C_{ij}\left\langle{\cal K}_i\Pi(\bar QQ){\cal K}^\dagger_j\right\rangle,
\label{te.matching}\end{equation}
where the left hand side is Taylor expanded in $q$. Each factor of $q$
Fourier transforms into a factor of the covariant derivative ${\bf D}$ on
the right hand side. Since each matrix element on the right of
eq.\ (\ref{te.matching}) corresponds to an unique matrix element in
eq.\ (\ref{te.nrqcd}), the order upto which the Taylor expansion is to be
performed is determined by the scaling of the non-perturbative matrix
elements with $v$. Since we require a classification of operators by the
angular momentum, it turns out to be very convenient to use spherical
tensor methods. These were used in an earlier paper \cite{pol} and are
used more extensively here.

In this paper we evaluate the cross sections to order
$\alpha_{\scriptscriptstyle S}v^9$. The Taylor expansion order, $N+N'\le6$
is obtained by setting $d=9$ and $E_d=M_d=0$ in eq.\ (\ref{te.rule}).
Furthermore, since we examine the leading term in perturbation theory, the
perturbative projector has only one term---
\begin{equation}
   \Pi(\bar QQ)\;=\;|\bar QQ\rangle\langle\bar QQ|.
\label{te.qqbproj}\end{equation}
In agreement with \cite{bchen} we use the relativistic normalisation
\begin{equation}
   \langle Q(p,\xi)\bar Q(q,\eta) | Q(p',\xi')\bar Q(q',\eta')\rangle
     \;=\;4 E_p E_q (2\pi)^6\delta^3(p-p')\delta^3(q-q'),
\label{te.norm}\end{equation}
with the spinor normalisations $\xi^\dagger\xi=\eta^\dagger\eta=1$. Since
$E_p=E_q=\sqrt{m^2+q^2}$, expanding this in $q^2$
allows us to write the spinor bilinears in terms of transition operators
built out of the heavy quark field. For example,
\begin{equation}
   \xi^\dagger\eta\;=\;{1\over2m}\langle\bar QQ(q)|\psi^\dagger\chi|0\rangle
      -{1\over2m^3}\langle\bar QQ(q)|\psi^\dagger{\bf D\cdot D}\chi|0\rangle
      +\cdots
\label{te.expn}\end{equation}

Conventionally the coefficient functions and matrix elements in
eq.\ (\ref{te.nrqcd}) were written with a non-relativistic normalisation
of the hadron states. In the threshold expansion technique it is more
convenient to retain a relativistic normalisation similiar to that in
eq.\ (\ref{te.norm}). The result for the cross section is the same in
either case, since a change in the definition of the matrix element is
compensated by a change in the coefficient function. To leading order
in $q$, the matrix elements in the notation of \cite{bbl} have to be
multiplied by $4m$ to obtain those in the relativistic normalisation
\cite{bchen}. As higher orders the relation is more complicated. In
this paper, we shall work entirely with the latter.

\section{\label{me}The Matrix Elements}

To leading order in $\alpha_{\scriptscriptstyle S}$, the kinematics is
very simple. The momenta of the initial particles are $p_1$ and $p_2$.
We take $p_1$ to lie in the positive $z$-direction and $p_2$ to be
oppositely directed. The momentum of the meson, $P=p_1+p_2$. As a result,
$s=P^2=M^2$, where $M$ is the meson mass.

The 4-momenta of $Q$ and $\bar Q$ ($p$ and $\bar p$ respectively) are
written as
\begin{equation}
   p\;=\;{1\over2}P+L_j q^j\qquad{\rm and}\qquad
   \bar p\;=\;{1\over2}P-L_j q^j.
\label{me.momdef}\end{equation}
Note that $p^2=\bar p^2=m^2$, where $m$ is the mass of the heavy quark.
The space-like vector $q$ is always defined in the rest frame of the pair,
and $L^\mu_j$ boosts it to any frame. We shall use Greek indices for Lorentz
tensors and Latin indices for Euclidean 3-tensors.

The following relations are easy to prove---
\begin{equation}
 p_1\cdot L_j\;=\; -{M\over2}\hat z_j,\qquad{\rm and}\qquad
 p_2\cdot L_j\;=\;  {M\over2}\hat z_j,
\label{me.dotl}\end{equation}
where $\hat z$ is the unit 3-vector in the $z$-direction.
They are consistent with the identity $P\cdot L_j=0$. We shall
use the two identities \cite{bchen}
\begin{equation}
   L_j\cdot L_k\;=\;-\delta_{ij},\qquad{\rm and}\qquad
   M\epsilon^{ijk}L_{\sigma k}\;=\;\epsilon_{\mu\nu\rho\sigma}
       L^\mu_i L^\nu_j P^\rho.
\label{me.ident}\end{equation}
Other relations can be written down \cite{bchen}, but are not important for
our computations. Note our convention $\epsilon^{0123}=1$.

This technique also depends on the usual non-relativistic reduction of Dirac
spinors which gives rise to the identities
\begin{equation}\begin{array}{rl}
   \bar u(p)\gamma^\mu v(\bar p)\;=\;& L^\mu_j\left[M\xi^\dagger\sigma^j\eta
     -{\displaystyle4\over\displaystyle M+2m}\delta_{mn}q^j q^m
                   \xi^\dagger\sigma^n\eta\right],\\
   \bar u(p)\gamma^\mu\gamma^5 v(\bar p)\;=\;&
      {\displaystyle2m\over\displaystyle M}P^\mu\xi^\dagger\eta - 2i L^\mu_m
       \epsilon_{mnj}q^n\xi^\dagger\sigma^j\eta.
\end{array}\label{me.spinor}\end{equation}
Here $\xi$ and $\eta$ are Pauli spinors and $\sigma^j$ are the usual Pauli
matrices. $M$ ($M^2=4m^2+4q^2$) is the invariant mass of the $\bar QQ$ system.

We work in a class of ghost-free gauges called the planar gauges \cite{ddt}.
These are defined by the polarisation sum for gluons
\begin{equation}
 \sum_\lambda\epsilon_\mu^\lambda(p)\epsilon_\nu^{*\lambda}(p)\;=\;
   d_{\mu\nu}(p)\;=\;-g_{\mu\nu}+{1\over p\cdot V}
   \left(p_\mu V_\nu+p_\nu V_\mu\right).
\label{me.gauge}\end{equation}
The propagator for a gluon of momentum $p$ is then given by
$G_{\mu\nu}(p)=d_{\mu\nu}(p)/p^2$. The vector $V$ defines a gauge choice.
We write $V=c_1 p_1+c_2 p_2$, with $c_1/c_2\sim{\cal O}(1)$. We verify
that all results are gauge invariant by the explicit check that they do
not depend on the arbitrary coefficients $c_1$ and $c_2$.

\subsection{$\bar qq\to\bar QQ$}

The matrix element for the subprocess $\bar qq\to\bar QQ$ is very simple.
It is given exactly by the expression
\begin{equation}
   {\cal M}\;=\;-{ig^2\over M^2}
      \left[\bar v(p_2)\gamma_\mu T^a u(p_1)\right] L^\mu_j
    \left[M\xi^\dagger\sigma^jT^a\eta - 
        {4\over M+2m}
      \,q^j\xi^\dagger(q\cdot\sigma)T^a\eta\right],
\label{me.qqmat}\end{equation}
where $u$ and $v$ are the light quark spinors. The equations of motion
for the initial state quarks has been used to obtain the explicitly gauge
invariant matrix element in eq.\ (\ref{me.qqmat}). The desired Taylor series
expansion is obtained by using the relation $M^2=4(m^2+q^2)$ to expand all
factors with $M$. Converting to spherical tensors \cite{rose}, we find
\begin{equation}
   q^2\;=\;-\sqrt3 [qq]^0_0\qquad{\rm and}\qquad
   \sigma\cdot q\;=\;-\sqrt3 [\sigma q]^0_0,
\label{me.spht}\end{equation}
where the notation $[\cdots]^J_M$ denotes a coupling to angular momentum
$J$ and helicity $M$ of the spherical tensors inside the square
brackets\footnote{The factor of $-\sqrt3$ in the
conversion of dot products was used earlier in eq.\ (\ref{pc.exam})}.
The remaining Euclidean vectors are converted to spherical tensors after
squaring the matrix element.

\subsection{$gg\to\bar QQ$}

The $s$-channel gluon exchange diagram can easily be reduced to the form
\begin{equation}
   {\cal M}_s\;=\;2g^2f_{abc}
      \left(A_j + {1\over2} \epsilon_1\cdot\epsilon_2\hat z_j\right)
      \left(\xi^\dagger\sigma^jT^c\eta - {4\over M(M+2m)}
           q^j \xi^\dagger q_i\sigma^iT^c\eta\right).
\label{me.schnl}\end{equation}
Here $\epsilon_i$ is the polarisation vector for the initial gluon of momentum
$p_i$, and $T^c$ is a colour generator. For convenience we have used the
notation
\begin{equation}
   A_i\;=\;{1\over M}\left(\epsilon_1\cdot L_i \epsilon_2\cdot p_1
       -\epsilon_2\cdot L_i \epsilon_1\cdot p_2\right).
\label{me.not1}\end{equation}

The $t$ and $u$ channel matrix elements require a little more work. The
colour factors can be reduced using the identity
\begin{equation}
   T_a T_b \;=\; {1\over6}\delta_{ab}+{1\over2}d_{abc}T^c
      +{i\over2}f_{abc}T^c.
\label{me.col}\end{equation}
The matrix elements can be written in the form
\begin{equation}
   {\cal M}_t\;=\;{t\over2p\cdot p_1}\qquad{\rm and}\qquad
   {\cal M}_u\;=\;{u\over2p\cdot p_2}.
\label{me.not2}\end{equation}
Then the factors of $1/p\cdot p_1$ and $1/p\cdot p_2$ permit a binomial
expansion in powers of $q$. The resulting series in $q$ has coefficients
which are simply related to $t+u$ and $t-u$. It is easy to check that
\begin{equation}\begin{array}{rl}
   t+u\;&=\;-\,\left({\displaystyle 4im\over\displaystyle M}\right)
      \epsilon_{\lambda\sigma\mu\nu}
         p_1^\lambda p_2^\sigma\epsilon_1^\mu\epsilon_2^\nu
                   \,(\xi^\dagger{\cal T}\eta)\\
     &\qquad\qquad\qquad
     -2M(A_j\hat z_m-A_m\hat z_j+B_{jm})
                   \,(q^m\xi^\dagger\sigma^j{\cal T}\eta)\\
     &\qquad\qquad
     +\left({\displaystyle8\over\displaystyle M+2m}\right)
         \delta_{jm}B_{np}\,(q^mq^nq^p\xi^\dagger\sigma^j{\cal T}\eta),\\
   t-u\;&=\;-\,2M^2(A_j+{1\over2}\epsilon_1\cdot\epsilon_2\hat z_j)
        \,(\xi^\dagger\sigma^j{\cal T}\eta)\\
     &\qquad\qquad
     +\left({\displaystyle8M\over\displaystyle M+2m}\right)
          \delta_{jm}(A_n+{1\over2}\epsilon_1\cdot\epsilon_2\hat z_n)
        \,(q^mq^n\xi^\dagger\sigma^j{\cal T}\eta).
\end{array}\label{me.tuchnl}\end{equation}
Here ${\cal T}$ stands either for the identity or a generator in the colour
$SU(3)$ space, depending on which part of the colour structure of eq.\ 
(\ref{me.col}) we consider. We have introduced the additional notation
\begin{equation}
   B_{ij}\;=\;\epsilon_1\cdot L_i \epsilon_2\cdot L_j
       +\epsilon_2\cdot L_i \epsilon_1\cdot L_j.
\label{me.not3}\end{equation}
Note that the binomial expansion is not the desired Taylor series expansion
in $q$, since both $t+u$ and $t-u$ involve $M$, which in turn depends on $q$.
However, it is an useful intermediate step, since it allows us to organise
the terms neatly.

The full matrix element can be written as
\begin{equation}
   {\cal M}\;=\;{1\over6}g^2\delta_{ab}S
               +{1\over2}g^2d_{abc}D^c
               +{i\over2}g^2f_{abc}F^c.
\label{me.ampl}\end{equation}
The colour amplitudes $S$ and $D$ involve only ${\cal M}_t+{\cal M}_u$,
whereas $F$ involves ${\cal M}_s$ as well as ${\cal M}_t-{\cal M}_u$. 

In order to write down our results, we find it convenient to introduce the
notation
\begin{equation}
  {\cal A}\;=\;{1\over M^2}\varepsilon_{\lambda\sigma\mu\nu}
         p_1^\lambda p_2^\sigma\epsilon_1^\mu\epsilon_2^\nu
     \qquad{\rm and}\qquad
  {\cal S}_{ij}\;=\;A_i\hat z_j+A_j\hat z_i-B_{ij}
               +\epsilon_1\cdot\epsilon_2\hat z_i\hat z_j.
\label{me.not4}\end{equation}
In order to identify all terms to order $v^9$ we need the
colour amplitude $S$ to order $q^5$---
\begin{equation}\begin{array}{rl}
   S\;&=\;-\left({\displaystyle8im\over\displaystyle M}\right){\cal A}
                   \,(\xi^\dagger\eta)
       + {\displaystyle 4\over\displaystyle M} {\cal S}_{jm}
                   \,(q^m\xi^\dagger\sigma^j\eta)
       - \left({\displaystyle32im\over\displaystyle M^3}\right){\cal A}
                   \hat z_m\hat z_n\,(q^mq^n\xi^\dagger\eta)
     \\ & \qquad\quad
       + {\displaystyle16\over\displaystyle M^3}
           \left[{\cal S}_{jm}\hat z_n\hat z_p
                     -{\displaystyle M\over\displaystyle M+2m}\delta_{jm}
                {\cal S}_{np}\right]\,(q^mq^nq^p\xi^\dagger\sigma^j\eta)
     \\ & \qquad\qquad
       - \left({\displaystyle128im\over\displaystyle M^5}\right){\cal A}
                   \hat z_m\hat z_n\hat z_p\hat z_r
             \,(q^mq^nq^pq^r\xi^\dagger\eta)
     \\ & \quad\quad
       + {\displaystyle64\over\displaystyle M^5}
           \left[{\cal S}_{jm}\hat z_n\hat z_p
                     -{\displaystyle M\over\displaystyle M+2m}\delta_{jm}
                {\cal S}_{np}\right]z_rz_s
                     \,(q^mq^nq^pq^rq^s\xi^\dagger\sigma^j\eta)
\end{array}\label{me.ampls}\end{equation}
To all orders, even powers of $q$ come with the tensor ${\cal A}$ and odd
powers with ${\cal S}$.
The amplitude $D$ differs only through having colour octet matrix elements
in place of the colour singlet ones shown above. For the colour amplitude
$F$ we need the expansion
\begin{equation}
   F^c\;=\;-\left({\displaystyle16im\over\displaystyle M^2}\right){\cal A}
                   \hat z_m\,(q^m\xi^\dagger T^c\eta)
       + {\displaystyle8\over\displaystyle M^2}{\cal S}_{jm}\hat z_n
                   \,(q^mq^n\xi^\dagger\sigma^jT^c\eta)
\label{me.amplf}\end{equation}
In this amplitude odd powers of $q$ come with the tensor ${\cal A}$ and even
powers with ${\cal S}$. In all three colour amplitudes, the terms in ${\cal A}$
are spin singlet and those in ${\cal S}$ are spin triplet.

The decomposition into spherical tensors can be performed partially at
this stage by using the identities
\begin{equation}
   \hat z_mq^m\;=\;[q]^1_0\qquad{\rm and}\qquad
   \hat z_m\hat z_nq^m q^n\;=\;\sqrt{2\over3}[qq]^2_0-\sqrt{1\over3}[qq]^0_0.
\label{me.spten}\end{equation}
The Euclidean indices on $\cal S$ are most conveniently converted after
squaring the matrix element.

A recurrence relation for the $i$-th term, $t_i$, in either of eqs.\ 
(\ref{me.ampls}) or (\ref{me.amplf}) is easy to write. We find that
\begin{equation}
   t_i\;=\;{4\over M^2}\hat z_a\hat z_b q^a q^b t_{i-2}.
\label{me.recur}\end{equation}
This holds for all $i>3$ in the $S$ and $D$ amplitudes and $i>4$ for
the $F$ amplitude. Also, for the $F$ amplitude this holds for $i=3$.
The $i=4$ term in $F$ is $(2/M)\hat z_nq^n$ times the $i=4$ term in
$D$. The required Taylor series expansion is then obtained by
expanding all factors containing $M$. This procedure is completely
systematic and may be performed, for example, by a Mathematica
program.

\subsection{$\gamma g\to\bar QQ$ and $\gamma\gamma\to\bar QQ$}

The matrix elements for the two processes $\gamma p\to\bar QQ$ and
$\gamma\gamma\to\bar QQ$ are closely related to the $gg$ amplitudes.
It is easy to check that
\begin{equation}
   {\cal M}_{\gamma g}\;=\; ge D,\qquad{\rm and}\qquad
   {\cal M}_{\gamma\gamma}\;=\; e^2 S,
\label{me.gamma}\end{equation}
where $D$ and $S$ are the colour amplitudes given in eq.\ (\ref{me.ampl}),
and $e$ is the charge of the heavy quark.

\section{The Cross Sections\label{cs}}

\subsection{$\bar qq\to\bar QQ$}

The squared matrix element for this process is easy to write down. After
summing over initial state helicities, the amplitude square can be expressed
in terms of matrix elements, over heavy-quark spinors, of products of $\sigma$
and $q$. At this stage a perturbative projector (eq.\ \ref{te.qqbproj}) is
introduced between the spinor bilinears in order to project on to $\bar QQ$
final states. The normalisation of these states involve the energies of the
quarks (see eq.\ \ref{te.norm}), and can be expanded in $q^2$, as shown in
eq.\ (\ref{te.expn}). The computation is complete once this expansion is
performed and the extra factors of $q$ arising from this appropriately
absorbed into the matrix elements. The non-perturbative matrix elements
needed for the cross sections of various charmonium states are listed in
Table \ref{cs.matelemq}.

\begin{table}\begin{center}\begin{tabular}{|c|c|c|}
\hline
H & d & Matrix Elements \\
\hline
$\eta_c$ & 7 & ${\cal O}_8\left({}^3S_1^0\right)$ \\
\hline
$h_c$ & 9 &  ${\cal O}_8\left({}^3S_1^0\right)$ \\
\hline
$J/\psi$ & 7 &    ${\cal O}_8\left({}^3S_1^0\right)$\\
         & 9 &    ${\cal P}_8\left({}^3S_1^0,{}^3S_1^2\right)$,
                  ${\cal O}_8\left({}^3P_1^2\right)$ \\
\hline
$\chi_J$ & 5 &    ${\cal O}_8\left({}^3S_1^0\right)$\\
         & 7 &    ${\cal P}_8\left({}^3S_1^0,{}^3S_1^2\right)$\\
         & 9 &    ${\cal O}_8\left({}^3S_1^2\right)$,
                  ${\cal P}_8\left({}^3S_1^0,{}^3S_1^4\right)$,
                  ${\cal O}_8\left({}^3D_1^2\right)$\\
\hline
\end{tabular}\end{center}
\caption[dummy]{The matrix elements from the $\bar qq$ process contributing
  to the cross section for all charmonium states $H$ at order $v^d$. We use
  $h_c$ as shorthand for the ${}^1P_1$ meson.}
\label{cs.matelemq}\end{table}

Finally, we list the parton level cross sections---
\begin{equation}\begin{array}{rl}
   \hat\sigma^{\eta_c}_{\bar qq} \;&=\;
       \displaystyle{\pi^3\alpha_s^2\over54m^4}
          \delta(\hat s-4m^2){\cal O}^{\eta_c}_8({}^3 S^0_1)\\

   \hat\sigma^{h_c}_{\bar qq} \;&=\;
       \displaystyle{\pi^3\alpha_s^2\over54m^4}
          \delta(\hat s-4m^2){\cal O}^{h_c}_8({}^3 S^0_1)\\

   \hat\sigma^{J/\psi}_{\bar qq} \;&=\;
       \displaystyle{\pi^3\alpha_s^2\over54 m^4}
          \delta(\hat s-4m^2)\biggl[{\cal O}^{J/\psi}_8({}^3 S^0_1)
       \\&\qquad\qquad
                +\displaystyle{1\over m^2}\biggl\{
                   \displaystyle{2\over\sqrt3}
                      {\cal P}^{J/\psi}_8({}^3 S^0_1,{}^3 S^2_1)
                  +\displaystyle{1\over4}
                      {\cal O}^{J/\psi}_8({}^3 P^2_1)\biggr\}\biggr]\\

   \hat\sigma^{\chi_J}_{\bar qq} \;&=\;
       \displaystyle{\pi^3\alpha_s^2\over54m^4}
          \delta(\hat s-4m^2)\biggl[{\cal O}^{\chi_J}_8({}^3 S^0_1)
                +\displaystyle{2\over\sqrt3m^2}
                     {\cal P}^{\chi_J}_8({}^3 S^0_1,{}^3 S^2_1)
       \\&\qquad
            +\displaystyle{1\over m^4}\left\{
                 \displaystyle{4\over3}{\cal O}^{\chi_J}_8({}^3 S^2_1)
                +\displaystyle{5\over12}{\cal O}^{\chi_J}_8({}^3 D^2_1)
                +\displaystyle{7\sqrt5\over12}
                     {\cal P}^{\chi_J}_8 ({}^3 S^0_1,{}^3 S^4_1)\right\}
          \biggr]
\end{array}\label{cs.xsecq}\end{equation}
See the appendix for details of the angular momentum coupling scheme used
in this paper.

Note that in any application to hadronic collisions, the parton level
centre of mass energy will be $\hat s=x_1 x_2 s$, where $s$ is the CM
energy of the hadrons and $x_1$ and $x_2$ are the momentum fractions
of the two partons. The contribution of this sub-process to the hadronic
cross section is then obtained by convoluting the above cross sections
with the appropriate parton density functions.

\subsection{$gg\to\bar QQ$}

The squared matrix element for the $gg$ process is more complicated,
but the extraction of the cross section follows exactly the same
steps as for the $\bar qq$ process. Denoting the average over initial
states of the product $\cal S^*$ by ${\cal S}\cdot {\cal S}^*$, we find
\begin{equation}
   {\cal S\cdot S^*}\;=\;\overline{\sum_{pol}}S_{jm}S^*_{j'm'}\;=\;
     \sum_{\lambda=\pm2}[\sigma q]^2_\lambda
                        [\sigma^\dagger q^\dagger]^2_{-\lambda}
    +{3\over2}[\sigma q]^0_0[\sigma^\dagger q^\dagger]^0_0.
\label{cs.ident}\end{equation}
Although $\sigma$ and $q$ are self-adjoint, we have retained the more
cumbersome notation in order to clarify the coupling of the different
angular momenta. Also, ${\cal A\cdot A^*}=1/8$ and ${\cal A\cdot S^*}=0$.
Consequently, even and odd terms in the three colour amplitudes of
eq.\ (\ref{me.ampl}) do not interfere with each other.

\begin{table}\begin{center}\begin{tabular}{|c|c|c|c|c|}
\hline
H & d & \multicolumn{3}{c|}{Colour amplitude} \\
       \cline{3-5} & & S & D & F \\ 
\hline
$\eta_c$ & 3 & ${\cal O}_1\left({}^1S_0^0\right)$ & & \\
         & 5 & ${\cal P}_1\left({}^1S_0^0,{}^1S_0^2\right)$ & & \\
         & 7 & ${\cal O}_1\left({}^1S_0^2\right)$,
		${\cal P}_1\left({}^1S_0^0,{}^1S^4_0\right)$  &
                   ${\cal O}_8\left({}^1S_0^0\right)$ &
                   ${\cal O}_8\left({}^1P_1^1\right)$ \\
\hline
$h_c$ & 5 &  & ${\cal O}_8\left({}^1S_0^0\right)$ & \\
      & 7 &  & ${\cal P}_8\left({}^1S_0^0,{}^1S_0^2\right)$ & \\
      & 9 &    ${\cal O}_1\left({}^1S_0^0\right)$ &
               ${\cal O}_8\left({}^1S_0^2\right)$,
               ${\cal O}_8\left({}^3P_J^1\right)$,&
               ${\cal O}_8\left({}^1P_1^1\right)$ \\
     &    &  & ${\cal O}_8\left({}^1D_2^2\right)$,
               ${\cal P}_8\left({}^1S_0^0,{}^1S_0^4\right)$&\\ 
\hline
$J/\psi$ & 7 & &  ${\cal O}_8\left({}^1S_0^0\right)$,
                  ${\cal O}_8\left({}^3P_J^1\right)$ & \\
         & 9 & &  ${\cal P}_8\left({}^1S_0^0,{}^1S_0^2\right)$,
                  ${\cal P}_8\left({}^3P_J^1,{}^3P_J^3\right)$ &
                  ${\cal O}_8\left({}^3P_{J^\prime}^2\right)$ \\
\hline
\end{tabular}\end{center}
\caption[dummy]{The matrix elements contributing to the cross section for
  some charmonium states $H$ at order $v^d$. We use $h_c$ as shorthand for
  the ${}^1P_1$ meson. $J=0,2$ and $J^\prime =1,2$}
\label{cs.matelem1}\end{table}

\begin{table}\begin{center}\begin{tabular}{|c|c|c|c|c|}
\hline
H & d & \multicolumn{3}{c|}{Colour amplitude} \\
       \cline{3-5} & & S & D & F \\ 
\hline
$\chi_0$ & 5 & ${\cal O}_1\left({}^3P_0^1\right)$ & & \\
         & 7 & ${\cal P}_1\left({}^3P_0^1,{}^3P_0^3\right)$ & & \\
         & 9 & ${\cal O}_1\left({}^3P_2^1\right)$,
               ${\cal O}_1\left({}^3P_0^3\right)$,&
               ${\cal O}_8\left({}^1S_0^0\right)$,&
               ${\cal O}_8\left({}^1P_1^1\right)$,
               ${\cal O}_8\left({}^3S_1^2\right)$, \\
         &   & ${\cal P}_1\left({}^3P_0^1,{}^3P_0^5\right)$ &
               ${\cal O}_8\left({}^3P_{0,2}^1\right)$ &
               ${\cal O}_8\left({}^3D_1^2\right)$\\
\hline
$\chi_1$ & 9 & ${\cal O}_1\left({}^3P_{0,2}^1\right)$ &
               ${\cal O}_8\left({}^1S_0^0\right)$, &
               ${\cal O}_8\left({}^1P_1^1\right)$,
               ${\cal O}_8\left({}^3S_1^2\right)$, \\
         &   &&${\cal O}_8\left({}^3P_{0,2}^1\right)$ &
               ${\cal O}_8\left({}^3D_{1,2}^2\right)$\\
\hline
$\chi_2$ & 5 & ${\cal O}_1\left({}^3P_2^1\right)$ & & \\
         & 7 & ${\cal P}_1\left({}^3P_2^1,{}^3P_2^3\right)$ & & \\
         & 9 & ${\cal O}_1\left({}^3P_0^1\right)$,
               ${\cal O}_1\left({}^3P_2^3\right)$, &
               ${\cal O}_8\left({}^1S_0^0\right)$,&
               ${\cal O}_8\left({}^1P_1^1\right)$,
               ${\cal O}_8\left({}^3S_1^2\right)$, \\
         &   & ${\cal P}_1\left({}^3P_2^1,{}^3P_2^5\right)$ &
               ${\cal O}_8\left({}^3P_{0,2}^1\right)$ &
               ${\cal O}_8\left({}^3D_{1,2,3}^2\right)$\\
\hline
\end{tabular}\end{center}
\caption[dummy]{The matrix elements contributing to the cross section for
  some charmonium states $H$ at order $v^d$.}
\label{cs.matelem2}\end{table}

In Tables \ref{cs.matelem1} and \ref{cs.matelem2} we list all the matrix
elements which appear in charmonium cross sections to order $v^9$. The
final parton level cross sections are listed in the following subsections.
The coefficients in the linear combinations of matrix elements appearing
there depend on the angular momentum coupling scheme. Our conventions are
set out in the appendix. The contributions of these processes to
hadronic cross sections are obtained by convoluting the sub-process cross
sections with appropriate parton densities.

\subsection{Direct $J/\psi$ cross section}
The direct $J/\psi$ subprocess cross section is
\begin{equation}
   \hat\sigma^{J/\psi}_{gg}(\hat s) \;=\;
          \varphi \left[{5\over48}\Theta^{J/\psi}_D(7)
                  +\left\{{5\over48}\Theta^{J/\psi}_D(9)
                        +{3\over16}\Theta^{J/\psi}_F(9)\right\}
		\right]
\label{cs.jpsi}\end{equation}
where
\begin{equation}
\varphi = {\pi^3\alpha_s^2\over4m^2}\delta(\hat s-4 m^2),
\label{cs.varphi}\end{equation}
and $\Theta^{J/\psi}_a(n)$ denotes combinations of non-perturbative
matrix elements from the colour amplitude $a$ ($=S$, $D$ or $F$) at
order $v^n$. Using the notation explained in the appendix, these can
be written as
\begin{equation}\begin{array}{rl}
   \Theta^{J/\psi}_D(7)\;=\;&
        {\displaystyle{1\over2m^2}}{\cal O}^{J/\psi}_8({}^1S_0^0)
       +{\displaystyle{1\over2m^4}}\left[
               3{\cal O}^{J/\psi}_8({}^3P_0^1)
              +{\displaystyle{4\over5}}{\cal O}^{J/\psi}_8({}^3P_2^1)
                                   \right] \\

   \Theta^{J/\psi}_D(9)\;=\;&
        {\displaystyle{1\over\sqrt3m^4}}{\cal P}^{J/\psi}_8({}^1S_0^0,{}^1S_0^2)
       +{\displaystyle{1\over\sqrt{15}m^6}}\biggl[
           {\displaystyle{35\over4}}
                    {\cal P}^{J/\psi}_8({}^3P_0^1,{}^3P_0^3)
      \\&\qquad
          +2 {\cal P}^{J/\psi}_8({}^3P_2^1,{}^3P_2^3)
                                   \biggr] \\

   \Theta^{J/\psi}_F(9)\;=\;& \displaystyle{1\over2 m^6} 
               \left[{1\over3} {\cal O}^{J/\psi}_8 ({}^3P^2_1)-
             {2\over5} {\cal O}^{J/\psi}_8 ({}^3P^2_2) \right]\\
\end{array}\label{cs.jpsime}\end{equation}

Previous computations \cite{bchen} have considered the expansion only
to order $v^7$. Heavy-quark spin symmetry gives the relation
${\cal O}^{J/\psi}_8 ({}^3P^1_2)=5{\cal O}^{J/\psi}_8 ({}^3P^1_0)$,
upto corrections of order $v^2$. Then at order $v^7$ accuracy this
can be used for a further reduction of $\theta^{J/\psi}_D(7)$. Since
we consider the expansion to order $v^9$, we cannot use this relation.

\subsection{$\chi_0$ cross section}
The $\chi_0$ subprocess cross section is
\begin{equation}\begin{array}{rl}
   \hat\sigma^{\chi_0}_{gg}(\hat s) \;=&\;
	\varphi \biggl[\displaystyle{1\over18}\Theta^{\chi_0}_S(5)
                     +\displaystyle{1\over18}\Theta^{\chi_0}_S(7)
          \\&\qquad\qquad\qquad
                     +\left\{\displaystyle{1\over18}\Theta^{\chi_0}_S(9) 
                     +\displaystyle{5\over48}\Theta^{\chi_0}_D(9)
                     +\displaystyle{3\over16}\Theta^{\chi_0}_F(9)\right\}
             \biggr].
\end{array}\label{cs.chi0}\end{equation}
The results to order $v^5$ are known from previous computations. This
work identifies all the combinations of matrix elements at orders $v^7$
and $v^9$. Note that the order $v^7$ term involves an off-diagonal
operator ({\sl i.e.\/}, a $\cal P$ term), and can be found only after
expanding the colour amplitude $S$ to third order in $q$.

Using the notation explained in the appendix, the combinations of
non-perturbative matrix elements appearing in eq.~(\ref{cs.chi0}) are
\begin{equation}\begin{array}{rl}
   \Theta^{\chi_0}_S(5)\;=\;&
        {\displaystyle{3\over2m^4}}{\cal O}^{\chi_0}_1({}^3P_0^1)\\

   \Theta^{\chi_0}_S(7)\;=\;&
        {\displaystyle{7\sqrt5\over4\sqrt3m^6}}
                {\cal P}^{\chi_0}_1({}^3P_0^1,{}^3P_0^3)\\

   \Theta^{\chi_0}_S(9)\;=\;&
        {\displaystyle{2\over5m^4}}{\cal O}^{\chi_0}_1({}^3P_2^1)
       +{\displaystyle{1\over8m^8}}\biggl[
           {\displaystyle{245\over9}}{\cal O}^{\chi_0}_1({}^3P_0^3)
          +{\displaystyle{149\sqrt7\over10\sqrt3}}
               {\cal P}^{\chi_0}_1({}^3P_0^1,{}^3P_0^5)
                                  \biggr]\\

   \Theta^{\chi_0}_D(9)\;=\;&
        {\displaystyle{1\over2m^2}}{\cal O}^{\chi_0}_8({}^1S_0^0)
       +{\displaystyle{1\over2m^4}}\left[
               3{\cal O}^{\chi_0}_8({}^3P_0^1)
              +{\displaystyle{4\over5}}{\cal O}^{\chi_0}_8({}^3P_2^1)
                                   \right] \\

   \Theta^{\chi_0}_F(9)\;=\;&
           \displaystyle{1\over 6 m^4}{\cal O}^{\chi_0}_8 ({}^1P^1_1)
           +\displaystyle{1\over18m^6} \biggl[
                  {\cal O}^{\chi_0}_8 ({}^3S^2_1)
                 +5{\cal O}^{\chi_0}_8 ({}^3D^2_1)
					   \biggr].
\end{array}\label{cs.chi0me}\end{equation}

\subsection{$\chi_1$ cross section}
\begin{equation}
   \hat\sigma^{\chi_1}_{gg}(\hat s) \;=\;
		\varphi \left[{1\over18}
        \Theta^{\chi_1}_S(9) +{5\over48} \Theta^{\chi_1}_D(9) 
        +{3\over16} \Theta^{\chi_1}_F(9)
		     \right]
\label{cs.chi1}\end{equation}
where the combinations of non-perturbative matrix elements are,
in the notation explained in the appendix,
\begin{equation}\begin{array}{rl}
   \Theta^{\chi_1}_S(9)\;=\;&
       {\displaystyle{1\over2m^4}}\left[
               3{\cal O}^{\chi_1}_1({}^3P_0^1)
              +{\displaystyle{4\over5}}{\cal O}^{\chi_1}_1({}^3P_2^1)
                                   \right] \\

   \Theta^{\chi_1}_D(9)\;=\;&
        {\displaystyle{1\over2m^2}}{\cal O}^{\chi_1}_8({}^1S_0^0)
       +{\displaystyle{1\over2m^4}}\left[
               3{\cal O}^{\chi_1}_8({}^3P_0^1)
              +{\displaystyle{4\over5}}{\cal O}^{\chi_1}_8({}^3P_2^1)
                                   \right] \\

   \Theta^{\chi_1}_F(9)\;=\;& \displaystyle{1\over 6 m^4}  
        {\cal O}^{\chi_1}_8 ({}^1P^1_1) 
       +\displaystyle{1\over3m^6} \biggl[
        \displaystyle{1\over6}{\cal O}^{\chi_1}_8 ({}^3S^2_1)
      \\&\qquad
        +\displaystyle{5\over6}{\cal O}^{\chi_1}_8 ({}^3D^2_1) 
        -\displaystyle{1\over5}{\cal O}_8 ({}^3D^2_2) 
					   \biggr].
\end{array}\label{cs.chi1me}\end{equation}
The $\chi_1$ is produced first at order $v^9$. The large branching ratio
for the decay $\chi_1\to J/\psi$ makes this a phenomenologically important
term, and is the main motivation for this work.

\subsection{$\chi_2$ cross section}
\begin{equation}\begin{array}{rl}
   \hat\sigma^{\chi_2}_{gg}(\hat s) \;=&\;\varphi
       \biggr[\displaystyle{1\over18}\Theta^{\chi_2}_S(5)
             +\displaystyle{1\over18}\Theta^{\chi_2}_S(7)
         \\&\qquad\qquad\qquad
             +\left\{\displaystyle{1\over18}\Theta^{\chi_2}_S(9)
                    +\displaystyle{5\over48}\Theta^{\chi_2}_D(9) 
                    +\displaystyle{3\over16}\Theta^{\chi_2}_F(9)\right\}
		     \biggr]
\end{array}\label{cs.chi2}\end{equation}
where the combinations of non-perturbative matrix elements are
\begin{equation}\begin{array}{rl}
   \Theta^{\chi_2}_S(5)\;=\;&
        {\displaystyle{2\over5m^4}}{\cal O}^{\chi_2}_1({}^3P_2^1)\\

   \Theta^{\chi_2}_S(7)\;=\;&
      {\displaystyle{2\over\sqrt{15}m^6}}
                {\cal P}^{\chi_2}_1({}^3P_2^1,{}^3P_2^3)\\

   \Theta^{\chi_2}_S(9)\;=\;&
        {\displaystyle{3\over2m^4}}{\cal O}^{\chi_2}_1({}^3P_0^1)
       +{\displaystyle{1\over75m^8}}\biggl[
           {\displaystyle{262\over9}}{\cal O}^{\chi_2}_1({}^3P_2^3)
          +{\displaystyle{141\sqrt3\over2\sqrt7}}
                  {\cal P}^{\chi_2}_1({}^3P_2^1,{}^3P_2^5)
                                  \biggr]\\

   \Theta^{\chi_2}_D(9)\;=\;&
        {\displaystyle{1\over2m^2}}{\cal O}^{\chi_2}_8({}^1S_0^0)
       +{\displaystyle{1\over2m^4}}\left[
               3{\cal O}^{\chi_2}_8({}^3P_0^1)
              +{\displaystyle{4\over5}}{\cal O}^{\chi_2}_8({}^3P_2^1)
                                   \right] \\

   \Theta^{\chi_2}_F(9)\;=\;& \displaystyle{1\over 6 m^4} 
        {\cal O}^{\chi_2}_8 ({}^1P^1_1) 
        + \displaystyle{1\over3m^6} \biggl[
        \displaystyle{1\over6}{\cal O}^{\chi_2}_8 ({}^3S^2_1) 
        +\displaystyle{5\over6}{\cal O}^{\chi_2}_8 ({}^3D^2_1)
       \\&\qquad
        -\displaystyle{1\over5}{\cal O}^{\chi_2}_8 ({}^3D^2_2) 
        +\displaystyle{2\over7}{\cal O}^{\chi_2}_8 ({}^3D^2_3) 
					   \biggr].
\end{array}\label{cs.chi2me}\end{equation}
See the appendix for an explanation of the notation.

\subsection{$\eta_c$ cross section}
The production cross section for $\eta_c$ is---
\begin{equation}\begin{array}{rl}
   \hat\sigma^{\eta_c}_{gg}(\hat s)\;=&\;\varphi
       \biggl[\displaystyle{1\over18}\Theta^{\eta_c}_S(3)
             +\displaystyle{1\over18}\Theta^{\eta_c}_S(5)
       \\&\qquad\qquad\qquad
            +\left\{\displaystyle{1\over18}\Theta^{\eta_c}_S(7)
                   +\displaystyle{5\over48}\Theta^{\eta_c}_D(7) 
                   +\displaystyle{3\over16}\Theta^{\eta_c}_F(7)\right\}
       \biggr],
\end{array}\label{cs.etac}\end{equation}
where the notation in the appendix can be used to write the combinations
of non-perturbative matrix elements as
\begin{equation}\begin{array}{rl}
   \Theta^{\eta_c}_S(3)\;=\;&
        {\displaystyle{1\over2m^2}}{\cal O}^{\eta_c}_1({}^1S_0^0)\\

   \Theta^{\eta_c}_S(5)\;=\;&
      {\displaystyle{1\over\sqrt3m^4}}
                {\cal P}^{\eta_c}_1({}^1S_0^0,{}^1S_0^2)\\

   \Theta^{\eta_c}_S(7)\;=\;&
        {\displaystyle{1\over3m^6}}\left[
            2{\cal O}^{\eta_c}_1({}^1S_0^2)
           +\displaystyle{4\over\sqrt5}{\cal P}^{\eta_c}_1({}^1S_0^0,{}^1S_0^4)
                                   \right]\\

   \Theta^{\eta_c}_D(7)\;=\;&
        {\displaystyle{1\over2m^2}}{\cal O}^{\eta_c}_8({}^1S_0^0)\\

   \Theta^{\eta_c}_F(7)\;=\;& \displaystyle{1\over 6 m^4} 
                          {\cal O}^{\eta_c}_8 ({}^1P^1_1)\\
\end{array}\label{cs.etacme}\end{equation}
Note that the colour amplitude $D$ first enters the cross section at
order $v^7$. Hence, the almost-elastic cross section $\gamma p\to\eta_c$
starts at order $v^7$.

\subsection{$h_c$ cross section}
The production cross section for the ${}^1P_1$ charmonium state is---
\begin{equation}\begin{array}{rl}
   \hat\sigma^{h_c}_{gg}(\hat s) \;=&\;\varphi
       \biggl[\displaystyle{5\over48}\Theta^{h_c}_D(5)
             +\displaystyle{5\over48}\Theta^{h_c}_D(7) 
         \\&\qquad\qquad\qquad
            +\left\{\displaystyle{1\over18}\Theta^{h_c}_S(9) 
                   +\displaystyle{5\over48}\Theta^{h_c}_D(9)
                   +\displaystyle{3\over16}\Theta^{h_c}_F(9)\right\}
 		     \biggr],
\end{array}\label{cs.hc}\end{equation}
where the combinations of non-perturbative matrix elements can be
written, using the notation in the appendix, as
\begin{equation}\begin{array}{rl}
   \Theta^{h_c}_D(5)\;=\;&
        {\displaystyle{1\over2m^2}}{\cal O}^{h_c}_8({}^1S_0^0)\\

   \Theta^{h_c}_D(7)\;=\;&
      {\displaystyle{1\over\sqrt3m^4}}
                {\cal P}^{h_c}_8({}^1S_0^0,{}^1S_0^2)\\

   \Theta^{h_c}_S(9)\;=\;&
        {\displaystyle{1\over2m^2}}{\cal O}^{h_c}_1({}^1S_0^0)\\

   \Theta^{h_c}_D(9)\;=\;&
        {\displaystyle{1\over3m^6}}\left[
            2{\cal O}^{h_c}_8({}^1S_0^2)
           +\displaystyle{4\over\sqrt5}{\cal P}^{h_c}_8({}^1S_0^0,{}^1S_0^4)
                                   \right]
      \\&\quad
       +{\displaystyle{1\over2m^4}}\left[
               3{\cal O}^{h_c}_8({}^3P_0^1)
              +{\displaystyle{4\over5}}{\cal O}^{h_c}_8({}^3P_2^1)
                                   \right]
       +{\displaystyle{1\over15m^6}}{\cal O}^{h_c}_8({}^1D_2^2)\\

   \Theta^{h_c}_F(9)\;=\;& 
        \displaystyle{1\over 6m^4} {\cal O}^{h_c}_8 ({}^1P^1_1).
\end{array}\label{cs.hcme}\end{equation}
Interestingly, only the colour amplitude $D$ enters the cross section
upto order $v^7$. As a consequence, the cross sections for $pp\to h_c$
and $\gamma p\to h_c$ require the same combinations of matrix elements
upto an accuracy of about 10\%.

\subsection{$\gamma g\to\bar QQ$ and $\gamma\gamma\to\bar QQ$}

The $\gamma g$ cross sections for the production of any quarkonium state,
eqs.\ (\ref{cs.jpsi})--(\ref{cs.hc}), can be obtained from those for the
$gg$ process by the following prescription. Replace
$\alpha_{\scriptscriptstyle S}^2$ in $\varphi$ (eq.\ \ref{cs.varphi}) by
$\alpha\alpha_{\scriptscriptstyle S}$, delete the $\Theta_S$ and $\Theta_F$
terms, and replace the colour factor $5/48$ for the terms in $\Theta_D$ by 2.
This follows from eq.\ (\ref{me.gamma}).

Similarly, the $\gamma\gamma$ cross sections are obtained from the
prescription--- replace $\alpha_{\scriptscriptstyle S}^2$ in $\varphi$
(eq.\ \ref{cs.varphi}) by $\alpha^2$, delete the $\Theta_D$ and $\Theta_F$
terms in eqs.\ (\ref{cs.jpsi})--(\ref{cs.hc}), and replace the colour
factor $1/18$ for the terms in $\Theta_S$ by 16.

\section{Discussion\label{disc}}

The number of non-perturbative matrix elements appearing in the cross
sections in Section \ref{cs} is rather large. In hadron-hadron collisions
the influence of the $\bar qq$ amplitudes is small compared to that of
the $gg$ amplitudes, since the gluon luminosity is much larger. This
still leaves us with the problem of determining a large number of
matrix elements.

The matrix elements appearing through the colour amplitude $D$ can be
isolated in almost-elastic $\gamma p$ collisions. At HERA it is now
possible to separate the diffractive components of the cross section
\cite{hera}. With this separation, it becomes possible to measure these
matrix elements with greater accuracy than was possible in the past
\cite{fleming}.

The matrix elements arising from the colour amplitude $S$ can be measured
in $\gamma\gamma$ collisions produced in $e^+e^-$ colliders, provided the
photons are point-like. Unfortunately, cross sections for $\gamma\gamma$
collisions at the LEP are likely to be dominated by resolved-photon effects
\cite{resolv}. On the other hand, double-tagged events with low hadronic
energy at the LEP correspond to the scattering of two highly virtual photons,
$\gamma^*\gamma^*$. The cross sections for $\gamma^*\gamma^*\to
J/\psi$ (and other quarkonia) can also be computed in NRQCD. It is easy
to see that these cross sections are obtained from the colour singlet part
of the amplitudes in eq.\ (\ref{me.not2}), the only difference being in
the denominators of the propagators. Hence, all the matrix elements from
the colour amplitude $S$ which appear in Section \ref{cs} will also appear
in this case, albeit in different linear combinations. Furthermore, the
coefficients would depend on the virtuality of each photon, leaving us with
the possibility of independent measurements of each of these matrix elements.
The full computation for such $\gamma^*\gamma^*$ processes will be presented
elsewhere.

The colour amplitude $F$ cannot be separated in any process, since it arises
only through $gg$ initial states. The corresponding matrix elements may be
obtained from $pp$, $\pi p$, $\bar pp$ collisions, as well as in $\gamma\gamma$
collisions where the resolved photon processes dominate.

Although quantitative estimates of low transverse momentum quarkonium
cross sections involves the prior analysis of many such experiments, it
is possible to make some crude but interesting estimates.

The $gg\to\eta_c$ cross section (eq.\ \ref{cs.etac}) involves only the
colour amplitude $S$ to next-to-leading order, $v^5$. The colour
amplitude $D$ first enters at order $v^7$. As a result, the $\gamma p
\to\eta_c$ cross section must be much less than the $pp\to\eta_c$ cross
section. In contrast, only the colour amplitude $D$ enters the $gg\to h_c$
cross section upto the next-to-leading order, $v^7$. Thus the $pp\to h_c$
cross section can be predicted to better than 10\% accuracy if the $\gamma
p\to h_c$ cross section is known. If the ${}^1P_1$ charmonium state is
identified definitively \cite{sridhar}, then this fact could serve as an
empirical test of NRQCD factorisation.

Further qualitative arguments depend on a more detailed analysis of the
non-perturbative matrix elements. Any such matrix element $\cal O$, with
mass dimension $D$, may be written in QCD as
\begin{equation}
   {\cal O}\;=\;\Lambda^D f\left({m_f\over\Lambda}\right),
\label{dis.scl}\end{equation}
where $\Lambda$ is the QCD scale and $m_f$ are the masses of the quarks
of flavour $f$. In the chiral limit, the light quark masses may be taken
to vanish, and the corresponding arguments of $f$ are then zero. For the
matrix elements which appear in quarkonium cross sections, the dependence
on the heavy quark mass is factored into the coefficient functions. After
such a factorisation, the resulting matrix element for the production of
a quarkonium $H$ can be written as
\begin{equation}
   {\cal O}^{\scriptscriptstyle H}\;=\;\Lambda^D
       f^{\scriptscriptstyle H}(v),
\label{dis.sclp}\end{equation}
where $v$ is the dimensionless velocity which organises the NRQCD expansion.
In the limit $v\to0$,
$f^{\scriptscriptstyle H}\to c_{\scriptscriptstyle H}v^d$, where
$c_{\scriptscriptstyle H}$ is a constant and $d$ is the power counting
dimension obtained, for example, by eq.\ (\ref{te.rule}). Power counting
in NRQCD is reliable if and only if $c_{\scriptscriptstyle H}$ obtained
from different operators (but for the same $H$) are of similiar orders of
magnitude.

Heavy-quark spin symmetry gives rise to further restrictions on these
constants, $c_{\scriptscriptstyle H}$. The operator expectation values from
different hadrons may be related using this approximate symmetry. Then, for
the same operator, the values of the constant obtained with two different
hadrons with the same radial wave-function, must be also of similiar magnitude,
after removing certain Clebsch-Gordan coefficients.

Such arguments may be used to study the convergence of the NRQCD series
for various cross sections. In the approximation of eq.\ (\ref{dis.sclp}),
the coefficient function times the matrix element is a function of two
dimensionless variables, $v$ and $z=\Lambda^2/m^2$. With the assumption
that all the constants $c_{\scriptscriptstyle H}$ are of similiar order,
the convergence properties are simple, since $z\ll v^2$. As an example,
we take the cross section for $gg\to\eta_c$ (eq.\ \ref{cs.etac}). After
removing all common factors, including dimensional quantities, we find
\begin{equation}
   \hat\sigma^{\eta_c}_{gg} \;\sim\;
      \left[{1\over36}v^3 + {5\over96}v^7 + \cdots\right] z^2 + {\cal O}(z^4).
\label{dis.etac}\end{equation}
The first term comes from $\Theta^{\eta_c}_S(3)$ and the second from
$\Theta^{\eta_c}_D(7)$ in eq.\ (\ref{cs.etacme}). The remaining terms are
of higher order in $z$, and converge rapidly. The two coefficients in
eq.\ (\ref{dis.etac}) are of similiar order, and it is quite possible that
the series in $v^2$ may be convergent for small $v$. Nevertheless, it is
interesting to note that the main correction to the leading term comes, not
from the order $v^5$ terms, but from one of the order $v^7$ terms.

A similiar argument may be used to assert that the most important term in
the cross section for $gg\to\chi_{\scriptscriptstyle J}$ is the
$\Theta_D(9)$ term (eqs.\ \ref{cs.chi0me}, \ref{cs.chi1me} and
\ref{cs.chi2me}), since it has the operator with the lowest mass dimension.
In that case, heavy-quark spin symmetry may be used to show that
\begin{equation}
   \sigma(\chi_0):\sigma(\chi_1):\sigma(\chi_2)\;\approx\;1:3:5,
\label{dis.pred}\end{equation}
both in $pp$ and almost-elastic $\gamma p$ collisions. The measured value
of this ratio in $pp$ and $\pi p$ collisions is $0.62\pm0.18\pm0.09$
\cite{expt}. Corrections to eq.\ (\ref{dis.pred}) would come from the
breaking of the approximate heavy-quark symmetry at higher orders in $v$.
Consequently, the same ratios should also be observed for the corresponding
bottomonium states. Since the coefficients for quarkonia with unequal principal
quantum numbers cannot be compared, the prediction of the direct to total
$J/\psi$ cross sections is a more detailed dynamical question.

It is interesting to test such a scaling hypothesis against the matrix
elements which have already been extracted. Unfortunately, a definitive
test cannot yet be made, since the values quoted for the same matrix
elements vary widely \cite{cho2,spain}. We take the examples of the
two sets
\begin{equation}
  \langle{\cal O}^H_8({}^3S_1)\rangle\qquad{\rm and}\qquad
  {1\over3}\langle{\cal O}^H_8({}^1S_0)\rangle+
      {1\over m^2}\langle{\cal O}^H_8({}^3P_0)\rangle,
\label{dis.matel}\end{equation}
for $J/\psi$, $\psi'$, $\Upsilon(1S)$ and $\Upsilon(2S)$. The two $L=0$
matrix elements give a common coefficient $c_{\scriptscriptstyle H}=c_1$,
and the $L=1$ matrix element gives a different coefficient $c_2$. For
the same $H$, if the scaling ideas are correct, then $(c_1-c_2)/(c_1+c_2)$
should be approximately zero. We find that for the $\Upsilon$ states the
fitted numbers \cite{cho2} are within one standard deviation of this
result. The values for $\psi'$ \cite{cho2} are about two standard deviations
away, whereas the many different fits for the $J/\psi$ \cite{cho2,spain}
lie between five and two standard deviations from this expectation. Although
one is tempted the accept the scaling argument, it turns out that $c_2$
is systematically larger than $c_1$. This clearly calls for an inclusion
of higher order effects, both in $\alpha_{\scriptscriptstyle S}$ and
in $v$. The latter, specially, may give rise to unforeseen corrections
in view of the pattern of coefficient functions seen here.

We would like to thank V.\ Ravishankar for discussions.

\appendix
\section{The Coupling Scheme\label{coupl}}

We use spherical tensor techniques for the reduction of the
matrix elements. Since all these are constructed by multiple
couplings of Euclidean 3-vectors, we need the spherical
tensor components for any vector $V$---
\begin{equation}
   V_0=V_z,\qquad V_1=-{1\over\sqrt2}(V_x+iV_y),
   \qquad V_{-1}={1\over\sqrt2}(V_x-iV_y).
\label{app.vec}\end{equation}
The labels on the spherical tensor components denote helicity.
Couplings of spherical tensors require the usual Clebsch-Gordan
coefficients \cite{rose}. We denote coupled tensors by the
notation $[a,b]^J_M$, where $J$ is the rank of the coupled tensor
and $M$ is the helicity.

Next we consider the complete specification of the state ${\cal K}_i|0
\rangle$, where ${\cal K}_i$ is one of the operators in eq.\ (\ref{te.nrqcd}).
When computing the NRQCD cross section to high orders in $v$, we may have
to couple a large number, $N$, of $q$'s, and, possibly, a Pauli matrix.
It is possible to write down the Clebsch-Gordan series for a reduction to
states of given $J$, $J_z$, $L$ and $S$, keeping explicitly the permutation
symmetry of the $q$'s. However, the coupling constants are not listed in the
literature, except for $N=3$. We choose instead to first ignore the permutation
symmetry, treat all the $q$'s as distinguishable operators, and use the usual
methods of angular momentum recouplings. When this is completed, we explicitly
symmetrise the results to get the desired expressions.

The total number of commuting operators required to specify
the state in the direct product basis is $2N+2S$. In the coupled basis
we have $N$ labels from each derivative operator, 2 from the total
angular momentum $J$ and the helicity $J_z$. If $S\ne0$ then two more
labels, $L$ and $S$ have to be specified. All these total to $N+2S+2$.
Consequently another $N-2$ labels have to be given for a complete
specification of the state. It is sufficient for this purpose to fix an
order of coupling the individual derivatives and the Pauli matrix and
specify the tensor rank at each coupling \cite{encyc}.

As an illustration we take the case $S=1$ and $N=4$. We specify a
coupling order either by the bracketed expression on the left or the
binary tree on the right---
\begin{equation}
\left[ \sigma \left[ \left[q q\right]^{j_1}
                     \left[q q\right]^{j_2} \right]^{j_3}
       \right]^J_{J_z}
    \;=\;\left.\qquad\right.
   \includegraphics{fig1.ps}
\label{app.exam}\end{equation}
Each node in the tree denotes a tensor whose angular momentum is to be
specified.
The root node (at the bottom) is the total angular momentum $J$. The
helicity, $J_z$, is also specified at this node. All the leaf nodes
(at the top) correspond to the individual Euclidean vectors being
coupled. A circled node represents a Pauli matrix. A simple counting
shows that all the required labels for the state are specified by this
means.

In the Taylor expansion of the appropriate matrix elements (section
\ref{me}) there is a natural pairing of derivatives. The factors of
$\hat z_m\hat z_n$ in the recurrence relation eq.\ (\ref{me.recur})
naturally pair these operators. In addition, $\delta_{mn}$ from $q^2$
factors or the tensor ${\cal S}_{np}$ in the fourth term of eq.\ 
(\ref{me.ampls}) also pair derivatives. The tensor ${\cal S}_{jm}$
pairs a Pauli matrix with a derivative. In addition, there is an unpaired
derivative in the $\bar qq$ amplitude (eq.\ \ref{me.qqmat}) and the
colour amplitude $F$ (eq.\ \ref{me.amplf}). A natural set of groupings
is then given by the following scheme---
\begin{enumerate}
   \item The spin singlet terms in the colour amplitudes $S$ and $D$ have
      even number of derivatives. These are grouped as
      \begin{equation}
         \biggl[[qq]\cdots[qq]\biggr] \;=\; \left.\qquad\right.
            \includegraphics{fig2.ps}
      \label{app.sch1}\end{equation}
   \item The spin singlet terms in the colour amplitude $F$ have
      an odd number of derivatives. Exactly one is unpaired. The rest are
      first coupled, and this unpaired derivative is coupled at the end---
      \begin{equation}
         \biggl[q\biggl[[qq]\cdots[qq]\biggr]\biggr]
                            \;=\; \left.\qquad\right.
            \includegraphics{fig4.ps}
      \label{app.sch3}\end{equation}
   \item The spin triplet terms in the colour amplitude $F$ or the $\bar qq$
      amplitude have an even number of derivatives. Exactly one is unpaired,
      and one paired with the Pauli matrix. The $[qq]$ pairs are first coupled;
      the unpaired derivative is coupled to this, and finally the $[\sigma q]$
      pair is recoupled to the result using a $6-J$ symbol. A further recoupling
      with another $6-J$ symbol then joins the two uncoupled derivatives---
      \begin{equation}
         \biggl[\sigma\biggl[q\biggl[q\biggl[[qq]\cdots[qq]\biggr]\biggr]
                     \biggr]\biggr] \to
         \biggl[\sigma\biggl[[qq]\biggl[[qq]\cdots[qq]\biggr]\biggr]\biggr]
                            \;=\; \left.\qquad\right.
            \includegraphics{fig5.ps}
      \label{app.sch4}\end{equation}
   \item The spin triplet terms in the colour amplitudes $S$ and $D$ have
      an odd number of derivatives. One is paired with a Pauli matrix. The
      rest of the derivatives are first coupled as before, and the remaining
      derivative is then recoupled to this. The coupling scheme is
      \begin{equation}
         \biggl[\sigma\biggl[q\biggl[[qq]\cdots[qq]\biggr]\biggr]\biggr]
                            \;=\; \left.\qquad\right.
            \includegraphics{fig3.ps}
      \label{app.sch2}\end{equation}
\end{enumerate}
In this work $N\le5$, and consequently, a choice of the order of couplings
of more than two pairs $[qq]$ is not required. At higher orders further
choices have to be made. Except at low orders ($N\le2$), the choice of
scheme is not unique. Different coupling schemes give rise to different
sets of basis states for ${\cal K}_i|0\rangle$. Unitary transformations can
always be found to relate these sets to each other. Although the linear
combinations of matrix elements appearing in the cross sections then look
different, they can be transformed into each other.

A possible phase ambiguity might remain in the definition of the states
because the coupling orders $[a,b]^j$ and $[b,a]^j$ differ by the phase
$(-1)^{j_a+j_b-j}$. Enumeration of all the cases arising in this problem
shows that there is no such ambiguity---
\begin{enumerate}
   \item Since $[qq]^j$ has only $j=0$ and 2, the phase is 1, and no
      ordering ambiguity exists.
   \item For $[[qq]^k_\lambda [qq]^j_0]^l_\lambda$, if $\lambda=0$, then
      we must have $k+j+l$ even. For integer $k$ and $j$ the phase is then
      1, and no ambiguity exists. When $\lambda=\pm2$, the phase is
      1 if $j=0$. If $j\ne0$, then the phase is not identically 1. However,
      in this case we always have a sum over helicities with multiplicative
      factors which are independent of $\lambda$. This projects out the terms
      with $k+j+l$ even, and removes the ambiguity.
   \item For $[[q]^1_0[[qq]\cdots[qq]]^l_\lambda]^j_\lambda$, the ordering
      is immaterial if $\lambda=0$, since $1+l+j$ is then even. When $\lambda
      =\pm2$, the sum over $\lambda$ removes the ambiguity.
   \item For $[\sigma q]^j$, since $j=0$ or 2, the phase is 1.
   \item For the final $L-S$ coupling, we always choose the Pauli matrix
      as the first factor.
\end{enumerate}
The binary tree is thus a complete and unambiguous specification of the
coupling scheme for our purposes.

After the recouplings are completed, we recognise that the tensors are
symmetric under interchange of all $q$'s. A symmetric tensor with $N$
3-d Euclidean indices reduces under the rotation group. The allowed angular
momenta are $L=N$, $N-2$, etc, down to 0 or 1, depending on whether $N$ is
even or odd. Simply counting the number of components in the tensor shows
that each allowed value of $L$ appears only once. As a result, it is
sufficient to label the symmetric states by $L$ and $N$ instead of all the
intermediate angular momenta.

The unitary transformation between the scheme specified by all the
intermediate angular momenta and the symmetric tensors can be carried out
explicitly. We find that---
\begin{equation}\begin{array}{rl}
  \left[\sigma\left[q [qq]^0\right]^1\right]^J_M\;=\;
      \left(\displaystyle{\sqrt5\over3}\right){}^3P_{\scriptscriptstyle J}^3
            \qquad&\qquad
  \left[\sigma\left[q [qq]^2\right]^1\right]^J_M\;=\;
      \left(\displaystyle{2\over3}\right){}^3P_{\scriptscriptstyle J}^3\\
  \left[[qq]^0[qq]^0\right]^0_0\;=\;
      \left(\displaystyle{\sqrt5\over3}\right){}^1{S}_0^4
            \qquad&\qquad
  \left[[qq]^2[qq]^2\right]^0_0\;=\;
      \left(\displaystyle{2\over3}\right){}^1{S}_0^4\\
  \left[\sigma\left[q\left[[qq]^0[qq]^0\right]^0\right]^1\right]^J_M\;=\;&
      \left(\displaystyle{1\over3}\sqrt{7\over3}\right)
                               {}^3P_{\scriptscriptstyle J}^5\\
  \left[\sigma\left[q\left[[qq]^2[qq]^2\right]^0\right]^1\right]^J_M\;=\;&
      \left(\displaystyle{2\over3}\sqrt{7\over15}\right)
                               {}^3P_{\scriptscriptstyle J}^5\\
  \left[\sigma\left[q\left[[qq]^0[qq]^2\right]^2\right]^1\right]^J_M\;=\;&
      \left(\displaystyle{2\over3}\sqrt{7\over15}\right)
                               {}^3P_{\scriptscriptstyle J}^5\\
  \left[\sigma\left[q\left[[qq]^2[qq]^2\right]^2\right]^1\right]^J_M\;=\;&
      \left(\displaystyle{4\over3\sqrt15}\right)
                               {}^3P_{\scriptscriptstyle J}^5\\
\end{array}\label{app.notation}\end{equation}

We specify the state ${\cal K}_i|0\rangle$ by a small extension of the
spectroscopic notation. In most cases we will write the state, or
equivalently, the operator, as ${}^{2S+1}L_J^N$, where $N$ gives the
number of derivatives. This allows us to use the new index $N$ for power
counting (see eq.\ \ref{te.rule}).

The matrix element $\langle{\cal K}_i\Pi(H){\cal K}_j\rangle$ is clearly
zero unless common quantum numbers of the two states ${\cal K}_i|0\rangle$
and ${\cal K}_j|0\rangle$ agree. Thus $L$, $S$, $J$ and $J_z$ must be
the same. Subject to these constraints, no quantum number reason prevents
a non-zero matrix element for bilinears with unequal $N$. Hence we retain
such off-diagonal operators in our computations.

\end{document}